\documentclass{svmult}

\usepackage{amsmath}
\usepackage{amssymb}
\usepackage{dsfont}

\usepackage[boxed, noend ]{algorithm2e}
\SetKwComment{refcomment}{[}{]}

\newcommand{\Card}{\mathrm{Card}}
\newcommand{\pl}{\mathrm{pl}}
\newcommand{\bel}{\mathrm{bel}}
\newcommand{\betp}{\mathrm{betP}}
\newcommand{\Bet}{\mathrm{Bet}}
\newcommand{\q}{\mathrm{q}}
\newcommand{\argmin}{\mathrm{argmin}}
\newcommand{\dist}{\mathrm{dist}}

\newcommand{\mC}{\mathcal{C}}
\newcommand{\mK}{\mathcal{K}}
\newcommand{\mO}{\mathcal{O}}

\title*{Controling the number of focal elements}
\subtitle{Some combinatorial considerations}
\author{Christophe {\sc Osswald}}

\begin{document}

\setcounter{minitocdepth}{2}

\dominitoc

\institute{Christophe {\sc Osswald} \at ENSTA Bretagne, Lab-STICC UMR 3192 \email Christophe.Osswald@ensta-bretagne.fr}

\maketitle

\abstract{A basic belief assignment can have up to $2^n$ focal
  elements, and combining them with a simple conjunctive operator will
  need $\mO(2^{2n})$ operations. This article proposes some techniques
  to limit the size of the focal sets of the bbas to be combined while
  preserving a large part of the information they carry. \\ The first
  section revisits some well-known definitions with an algorithmic
  point of vue. The second section proposes a matrix way of building
  the least committed isopignistic, and extends it to some other
  bodies of evidence. The third section adapts the $k$-means algorithm
  for an unsupervized clustering of the focal elements of a given bba.
}


\section{General considerations on basic belief assignments}

Let the finite set $X=\{x_1, \ldots, x_n\}$ be our frame of
discernment. The size of $X$ will be noted $n=|X|$. The set of all the
subsets of $X$ will be noted $2^X$.

\begin{definition} \cite{shafer76}
  The application $m$ from $2^X$ to $[0,1]$ is a {\em basic belief
    assignment} (bba) if~:
  \begin{equation}
    \sum_{A\subseteq X} m(A) = 1
  \end{equation}
\end{definition}

The constraint of {\em closed world} is modeled by
$m(\emptyset)=0$. If $m(\emptyset)$ is greater than 0, we either have
an {\em open world} or a {\em conflict} within the information.

\begin{definition} \label{def:focal}
  Let $m$ be a bba on $X$. $A\subseteq X$ is a {\em focal element} of $m$ if
  $m(A)>0$. The {\em focal set} of $m$ is composed of all its focal
  elements~: 
  \begin{equation}
    F(m) = \{A\subseteq X~|~m(A)>0\}
  \end{equation}
  The size of $m$ is noted $|m| = \Card(F(m))$. 
\end{definition}

Of course, $|m|\leqslant 2^n$. In most applications, $|m|$ will
be very small compared to $2^n$ when a bba is constructed from a
source's information, but after some steps of combination, this limit
can be reached.

\begin{definition}
  \label{def:belief}
  Let $m$ be a bba on $X$. The most usual {\em bodies of evidence} are~:
  \begin{itemize}
  \item The {\em belief}:\vspace{-2mm}
    \begin{equation}
      { 
        \renewcommand{\arraystretch}{0.8}
        \bel(A) = \sum_{\begin{array}{c}
            \scriptstyle B\subseteq A,\\
            \scriptstyle B\neq\emptyset
        \end{array}} m(B)  = \sum_{\begin{array}{c}
            \scriptstyle B\subseteq A,\\
            \scriptstyle B\neq\emptyset,\\
            \scriptstyle B \in F(m)
        \end{array}}  m(B)
      }
    \end{equation}
  \item The {\em plausibility}:\vspace{-2mm}
    \begin{equation}
      {         
        \renewcommand{\arraystretch}{0.8}
        \pl(A) = \sum_{B\cap A\neq \emptyset} m(B)  = \sum_{\begin{array}{c}
            \scriptstyle B\cap A\neq \emptyset,\\
            \scriptstyle B \in F(m)
        \end{array}}  m(B)
      }
    \end{equation}
  \item The {\em commonality}:\vspace{-2mm}
      \begin{equation}
        {         
          \renewcommand{\arraystretch}{0.8}
          \q(A) = \sum_{B\supseteq A} m(B)  = \sum_{\begin{array}{c}
              \scriptstyle B\supseteq A,\\
              \scriptstyle B \in F(m)
          \end{array}}  m(B)
        }
      \end{equation}
 \item The {\em pignisitic probability}, which is additive (knowing
    $\betp(\{x\})$ for all $x\in X$ is sufficient):\vspace{-2mm}
    \begin{equation}
      \hspace{-6mm}\betp(A) = { \renewcommand{\arraystretch}{0.8}
\frac{1}{1-m(\emptyset)}\!\!\!\sum_{B\subseteq X} \frac{|A\cap B|}{|B|}m(B)}
   =
 { \renewcommand{\arraystretch}{0.8}
        \frac{1}{1-m(\emptyset)}\!\!\!\!\sum_{\begin{array}{c}
            \scriptstyle B \in F(m) 
        \end{array}} \frac{|A\cap B|}{|B|}m(B) }
  \end{equation}
 
\end{itemize}
When the context is not obvious, the bba used to define the body of
evidence will be placed as an index~: $\betp_m(A)$ instead of
$\betp(m)$.
\end{definition}

In the definition \ref{def:belief}, the first expression concerns all
the subsets of $X$, and the second expression concerns only the focal
elements. Therefore, if $f$ is either of the bodies of evidence, and
$A$ a subset of $X$, a natural implementation of the equation brings
an algorithm which calculates $f(A)$ in $\mO(2^n)$ operations with the
first expression. As the second expression only browses the focal
set of $m$, its complexity is $\mO(|m|)$, for the same result. \\


The most popular combination operator is the non-normalized
conjunctive rule, also known as Smet's rule. It is a quite simple
operator to implement; it is associative, and therefore allows to
combine many sources.

\begin{definition}
  Let $m_1$ and $m_2$ be two bbas on $X$. The {\em conjunctive
    combination} of $m_1$ and $m_2$ is a bba on $X$, $m_1\oplus m_2$, defined
  by~:\vspace{-2mm}
  \begin{equation}
    \label{eq:conj}
    {
      \renewcommand{\arraystretch}{0.8}
      (m_1\oplus m_2)(A) = \sum_{\begin{array}{c}
          \scriptstyle B \subseteq X, \\
          \scriptstyle C \subseteq X, \\
          \scriptstyle B\cap C = A
        \end{array}
      } m_1(B) m_2(C) = \sum_{\begin{array}{c}
        \scriptstyle B \in F(m_1),\\
        \scriptstyle C \in F(m_2), \\
        \scriptstyle B \cap C = A
        \end{array}
      } m_1(B) m_2(C)
    }
  \end{equation}
\end{definition}

The cost for calculating $B\cap C$ is $\mO(n)$. The first expression
brings an algorithm in $\mO\left(n2^{2n}\right)$ operations for
calculating $(m_1\oplus m_2)(A)$, and $\mO\left(n2^{3n}\right)$ for
determining $m_1\oplus m_2$. The second expression brings an algorithm in
$\mO\left(n|m_1||m_2|\right)$ operations for calculating
$(m_1\oplus m_2)(A)=(m_1\oplus m_2)(A)$, and $\mO\left(n2^{n}|m_1||m_2|\right)$
for determining $m_1\oplus m_2$.


Smets \cite{smets02} proposed a nice implementation in $\mO(n2^n)$
operations for transformations between bba and commonality. The
conjunctive combination of the commonality functions is a simple
multiplication, which is linear, but on vectors having a size of
$2^n$.

The expression (\ref{eq:conj}), nor the commonality, can prevent us
from making operations on non-focal elements of $m_1\oplus m_2$. Let
the bba be implemented by an adaptive structure that contains
information only for its focal elements. A hashtable is a convenient
way for it. The algorithm \ref{alg:conj} uses only $\mO(n|m_1||m_2|)$
to build $m_1\oplus m_2$.

The size of $m_\cap$ is at most $|m_1||m_2|$. The algorithm coming
from (\ref{eq:conj}) needs to be executed for all the subsets of $X$,
but the algorithm \ref{alg:conj} only works on the focal elements of
$m_\cap$, and does not compute useless intersections
\cite{smets94}. Using a hashtable for the focal elements, with a
hashcode calculation in $\mO(n)$ operations, the conjunctive
combination takes $\mO(n|m_1||m_2|)$ operations.

\begin{algorithm}[H]
  \dontprintsemicolon
  \KwData{bbas $m_1$, $m_2$}
  \KwResult{bba $m_\cap$}
  \ForAll{$B\in m1$}{
    \ForAll{$C\in m_2$}{
      \eIf{$B\cap C \in m_\cap$}{
        $m_\cap(B\cap C) \gets m_\cap(B\cap C) + m_1(B)m_2(C)$
      }{Add $B\cap C$ to $m_\cap$\;
        $m_\cap(B\cap C) \gets m_1(B)m_2(C)$}}}
  \caption{Conjunctive combination}
  \label{alg:conj}
\end{algorithm}

However, the very nature of the combination operator brings a
combinatorial explosion of the focal set. Let $m_i$ be the bba defined
by $m_i(X)=\frac{1}{2}$ and $m_i(X\backslash\{x_i\})=\frac{1}{2}$:
$|m_i|=2$. Let $m_\cap$ be the conjunctive combination of all those
bbas~: $m_\cap = m_1 \oplus \ldots \oplus m_n$. For any $A\subseteq
X$, $m_\cap(A)=\frac{1}{2^n}$. Therefore, $F(m_\cap)=2^X$ and
$|m_\cap|=2^n$.
\\

The objective of the following sections will be to guarantee that the
size of a bba cannot be too large, and to respect its nature as much
as possible.

\section{Linear algebra for bbas}

The definition \ref{def:belief} builds the bodies of evidence $\bel$,
$\pl$, $\betp$ and $\q$ as linear transformations of $m$. Considering
a bba $m$ on $X$ and an integer $K$, our objective will be to build an
bba $m'$ on $X$ such that $|m'|\leqslant K$ and $f_{m'}(A) = f_{m}(A)$
for some bodies of evidence $f$ and some subsets $A$ of $X$.

Within this section, we forbid $\emptyset$ to be a focal element of
$m$, and we do not allow it to become a focal element of $m'$. As
convenient consequences, we have $\bel(A)\leqslant \betp(A)\leqslant
\pl(A)$, $\bel(X)=1$, and $\pl(X)=1$. \\

\label{ss:isopign}

A popular and efficient way to build a bba from a probability or
another source of uncertain information is to build a least committed
bba having the same pignistic probability than the source
\cite{smets90b}.

\begin{definition}
  \label{def:isopign}
  Let $m$ be a bba on $X$. A bba $m'$ is an {\em isopignisitic}
  of $m$ if 
  \begin{equation}
    \forall x\in X,~ \betp_m(x) = \betp_{m'}(x)
  \end{equation}
  The bba $m'$ is the {\em least committed isopignistic} of $m$ if
  for any isopignistic $m''$ of $m$ and for any $A\subseteq X$,
  $\pl_{m'}(A)\geqslant \pl_{m''}(A)$.
\end{definition}

The algorithm \ref{alg:isopign} builds the least committed isopignistic
in $\mO(n^2+n|m|)$ operations. It contains at most $n$ focal
elements. \\

\begin{algorithm}[H]
  \dontprintsemicolon
  \KwData{bba $m$ on $X$}
  \KwResult{bba $m'$ on $X$}
  \ForAll{$x\in X$}{
    Calculate $p[i] = \betp(x)$\;
  }
  $A \gets X$; $k \gets |X|$\;
  \While{$\max(p)\neq 0$}{
    $i \gets \argmin(p)$\;
    $m'(A) \gets kp[i]$\;
    \ForAll{$j \in p$}{
      $p[j] \gets p[j]-p[i]$\;
    }
    Delete element $i$ from $p$\;
    $A \gets A\backslash\{x_i\}$; $k\gets k-1$\;
  }
  \caption{Building the least committed isopignistic}
  \label{alg:isopign}
\end{algorithm}

If we calculate $\betp(x)$ for all $x\in X$, and order the elements of
$X$ such that $p_i = \betp(x_i)\geqslant \betp(x_{i+1}) = p_{i+1}$,
the focal elements of the least committed isopignistic are a subset of
the $A_i=\{x_1, \ldots, x_i\}$.

We have
\begin{equation}
  p_i = \betp(x_i) = \sum_{k=i}^n \frac{1}{k}m'(A_k)
\end{equation}

Let $p$ be the vector of the $p_i$ and $y$ be the vector of the
$m'(A_i)$. We have $p=\Bet y$ with $\Bet$ a $n\times n$ matrix, triangular
and inversible. Therefore $y=\Bet^{-1}p$, with
\begin{equation}
  \!\!\!\!\!\Bet = \begin{pmatrix}
    1 & \frac{1}{2} & \frac{1}{3} & \cdots & \frac{1}{n-1} & \frac{1}{n} \\
    0 & \frac{1}{2} & \frac{1}{3} & \cdots & \frac{1}{n-1} & \frac{1}{n} \\
    \vdots & 0  & \frac{1}{3} & \cdots & \frac{1}{n-1} & \frac{1}{n} \\
    \vdots & & \ddots & \ddots & & \vdots \\
    \vdots & & & \ddots & \frac{1}{n-1} & \vdots \\
    0 & \cdots & \cdots & \cdots & 0 & \frac{1}{n}
  \end{pmatrix},
  ~
  \Bet^{-1} = \begin{pmatrix}
    1 & -1 & 0 & \cdots & \cdots & 0 \\
    0 & 2 & -2 & 0 &  & \vdots \\
    0 & 0  & 3 & -3 & \ddots & \vdots \\
    \vdots &  & \ddots & \ddots & \ddots & 0 \\
    \vdots & & & \ddots & (n\!\!-\!\!1) & -(n\!\!-\!\!1) \\
    0 & \cdots & \cdots & \cdots & 0 & n
  \end{pmatrix}
\end{equation}

As $\Bet^{-1}$ is a triangular band matrix, we can compute all the
$m'(A_i)$ from $p_i$ in $\mO(n)$ operations.

With $\mO(n|m|)$ operations for computing $\betp$, $\mO(n\ln n)$
operations for sorting $X$, $\mO(n)$ operations for building the sets
$A_i$ (with an adapted data structure) and $\mO(n)$ operations for
solving the linear system, building the least committed isopignistic
costs $\mO(n(\ln n + |m|))$ operations. Usually, $|m|\gg \ln n$, and
the cost of the least committed isopignistic is not greater than the
cost of computing $\betp(x)$ for the elements of $X$. \\


The interval $[\bel(A), \pl(A)]$, containing $\betp(A)$, can be
interpretated as an uncertainty on $A$ \cite{janez96b}. For
singletons, $\bel$ is trivial: $\bel(x)=m(x)$. For sets of size $n-1$,
$\pl$ is trivial: $\pl(X\backslash\{x\})=1-m(\{x\})$. Considering the
non-trivial bodies of evidence on the sets of interest $\{x_1\}$,
\ldots, $\{x_n\}$, $B_1=X\backslash\{x_1\}$, \ldots,
$B_n=X\backslash\{x_n\}$, we search a bba $m'$ with those focal
elements, forming a vector
\begin{equation}
\mathbf{y} = \left(m'(\{x_1\}), \ldots, m'(\{x_n\}), m'(B_1), \ldots, m'(B_n)\right)^T
\end{equation}
which verifies:\vspace{-2mm}
\begin{eqnarray}
  \forall i\in\lfloor 1, n\rfloor, & & \pl_{m'}(\{x_i\}) = \pl_m(\{x_i\}) \\
  \forall i\in\lfloor 1, n\rfloor, & & \bel_{m'}(B_i) = \bel_m(B_i) 
\end{eqnarray}

We have: \vspace{-2mm}
\begin{eqnarray}
  \pl_{m'}(\{x_i\}) & = & m'(\{x_i\}) + \sum_{j\neq i} m'(B_j) \\
  \bel_{m'}(B_i) & = & \sum_{j\neq i} m'(\{x_j\}) + m'(B_i)
\end{eqnarray}

As $\forall i$, $\pl_{m'}(\{x_i\}) + \bel_{m'}(B_i) = \sum_i
m'(\{x_i\}) + \sum_i m'(B_j)$, there are only $n+1$ independent
equations among the $2n$ listed above: we cannot guarantee to kep
at the same time $\pl_m(\{x_i\})$ and $\bel_m(B_i)$ on those $2n$
focal elements. 

As $\q(B_i)=m(B_i)+m(X)$ and $\q(\{x_i\}) = \pl(\{x_i\})$, introducing
commonality does not bring any new independent equation.

\subsection{Mixing $\Bet$ with other bodies of evidence}
\label{ss:mix}

Here we search a bba with $2n$ focal elements which is an isopignistic
of $m$ and respects an other body of evidence on some focal elements.
In the following examples, we allow the $A_i$ obtained in section
\ref{ss:isopign} to be focal elements, and we complete them with
$(\{x_i\})_{i\in\lfloor 1, n\rfloor}$ or the $(B_i)_{i\in\lfloor 1,
  n\rfloor}$. \\

{\bf With plausibility}, we should use the focal elements
$(\{x_i\})_{i\in\lfloor 1, n\rfloor}$. We build a vector\vspace{-2mm}
\begin{equation}
  \mathbf{y} = \left(m'(\{x_1\}), \ldots, m'(\{x_n\}), m'(A_1), \ldots, m'(A_n)\right)^T
\end{equation}
The constraints are:\vspace{-2mm}
\begin{eqnarray}
  \betp(x_i) & = & m'(\{x_i\}) + \sum_{k=i}^n \frac{1}{k}m'(A_k) \\
  \pl(\{x_i\}) & = &  m'(\{x_i\}) + \sum_{k=i}^n m'(A_k) 
\end{eqnarray}
As $A_1=\{x_1\}$, we cannot have $m'(A_1)\neq m'(\{x_1\})$; we have
only $2n-1$ focal elements. We drop the term $m'(\{x_1\})$ in $y$, and
the constraint on $\pl(\{x_i\})$ to obtain a matrix $P$ such that
\begin{equation}
  P\mathbf{y} = (\pl_m(\{x_2\}), \ldots, \pl_m(\{x_2\}), \betp(x_1), \ldots,
  \betp(x_n))^T
\end{equation}
The matrix $P_4$ and more generally $P_n$ are:\vspace{-2mm}
\begin{equation}
  P_4 = \begin{pmatrix}
    0 & 0 & 0 & 1 & \frac{1}{2} & \frac{1}{3} & \frac{1}{4} \\
    1 & 0 & 0 & 0 & \frac{1}{2} & \frac{1}{3} & \frac{1}{4} \\
    0 & 1 & 0 & 0 & 0   & \frac{1}{3} & \frac{1}{4} \\
    0 & 0 & 1 & 0 & 0   & 0   & \frac{1}{4} \\
    1 & 0 & 0 & 0 & 1 & 1 & 1 \\ 
    0 & 1 & 0 & 0 & 0 & 1 & 1 \\ 
    0 & 0 & 1 & 0 & 0 & 0 & 1 \\ 
  \end{pmatrix}, ~~~~ P_n = \left(\begin{array}{c|c}
    \begin{array}{c}
      0 \cdots 0 \\
      \hline
      ~ \\
      I_{n-1} \\ ~
    \end{array} & \Bet_n \\
    \hline
    I_{n-1} & \begin{array}{c|c}
        0  & \\
        \vdots & U_{n-1} \\
        0 & 
    \end{array}
  \end{array}\right)
\end{equation}
where $\Bet_n$ is matrix obtained in the section \ref{ss:isopign} and
$U_{n-1}$ the upper triangular $(n\!-\!1)\!\times\!(n\!-\!1)$ matrix full of 1.

The matrix $P_n$ is inversible, and we can solve this system in
$\mO(n^3)$ operations. Overall, we can reduce the focal set of $m$ to
$2n-1$ focal elements in $\mO(n(n^2+|m|))$ operations, respecting
$\betp$ and $\pl$ on the singletons. \\

{\bf With commonality}, we obtain the same results~: $\q(\{x_i\})
= \pl(\{x_i\})$. \\

{\bf With belief}, we should use $(B_i)_{i\in\lfloor 1, n\rfloor}$ as
focal elements instead of $(\{x_i\})$. As $\bel(B_i)+\pl(\{x_i\})=1$,
we obtain another -- but similar -- $(2n\!-\!1)\!\times\!(2n\!-\!1)$
inversible matrix.

\section{Optimatization by $k$-means}

\cite{denoeux02} proposed to reduce a bba by
adapting the single linkage hierarchical clustering algorithm to
coarsen its focal set. Another interesting family of unsupervized
clustering algorithm are the $k$-means techniques, born from the
ISODATA method of 
\cite{kn_Bal_Hal_65}. One can adapt
this method to find a subset $\mK$ of $2^X$ limited in size:
$|\mK|\leqslant k$.

Usual $k$-means does not guarantee an optimal choice of centers:
finding them is equivalent to the {\sc minimum-$k$ center}, which is a
NP-Complete problem \cite{kn_Gar_Joh_79}. The convergence of the
$k$-means algorithm is guaranteed, but only to a local minimum
of the intra-cluster variance. \vspace{1mm}

\begin{algorithm}[H]
  \dontprintsemicolon
  \KwData{bba $m$, integer $k$ with $k\leqslant |m|$}
  \KwResult{bba $m_k$}
  Let $C[1]$, \ldots, $C[k]$ be $k$ focal elements of $m$\refcomment*[f]{1}\;
  \Repeat{ending condition reached}{
    \lForAll{$j\leqslant k$}{$\mC[j]\gets\emptyset$\;}
    \ForAll(\refcomment*[f]{2}){$A\in m$}{
      $\mC[\argmin(\dist(A, C_j))] \gets \mC[\argmin(\dist(A, C_j))]\cup\{A\}$\;
    }
    \lForAll(\refcomment*[f]{3}){$j\leqslant k$}{$C[j] \gets$ center of $\mC[j]$}}(\refcomment*[f]{4})
  \ForAll{$j\leqslant k$}{
    $m_k(C[j]) \gets \sum_{A\in \mC[j]} m(A)$
  }
  \caption{$k$-means, in a general way that applies to focal
    elements.}
  \label{alg:k-means}
\end{algorithm}
\begin{description}
\item[{\bf [1]}] It is natural to initialize the algorithm with the $k$
  focal elements with the greatest masses. But, as the algorithm
  converges -- if it converges -- to a local minimum, it should be a
  good idea to execute various instances, with random starting sets.
\item[{\bf [2]}] The focal element $A$ is affected to the center
  $C[j]$ such that\vspace{-2mm}
  \begin{equation}
    \dist(A,C[j])=\left|\left(A\cap\overline{C[j]}\right)\cup\left(\overline{A}\cap
      C[j]\right)\right|
  \end{equation}
  is minimal. It corresponds to a natural $L_1$ distance based on an
  exclusive {\sc or}. In case of equal distances to different centers,
  it is possible to:
  \begin{itemize}
  \item choose a random one \hfill {\em (the algorithm is no longer deterministic)}
  \item use a lexicographical order \hfill {\em (elements are no longer equivalent)}
  \item try to build balanced clusters \hfill {\em (the underlying problem is NP-complete)}
  \end{itemize}
\item[{\bf [3]}] The usual $k$-means technique uses the geometrical
  barycenter of the focal sets of $\mC[j]$ seen as points of
  $[0,1]^n$~: $C[j] \gets \sum_{A\in \mC[j]} m(A)A$.\\
  It would build fuzzy focal
  elements, which is not the way the definition \ref{def:focal}
  accepts them. Therefore, we put $x$ in the new $C[j]$ if
  and only if~:\vspace{-2mm}
  \begin{equation}
    \sum_{A\in \mC[j], x\in A} m(A) > \sum_{A\in \mC[j], x\not\in A} m(A)
  \end{equation}
\item[{\bf [4]}] As we ``move'' the centers of the classes to the
  nearest sharp subset of $X$, the total intra-cluster variance is not
  necessarily decreasing. Therefore, the ending condition must include
  a maximum steps number, and/or test the cycles it should encounter.
\end{description}

A step of the algorithm \ref{alg:k-means} costs $\mO(kn|m|)$
operations. A reasonable number of steps before ending the loop is
$k$, and we obtain an algorithm in $\mO(k^2n|m|)$ operations. If we
want to compare this approach with the ones of the section
\ref{ss:mix}, we should use $k=2n-1$, and get an algorithm in
$\mO(n^3|m|)$ operations.

\section{Conclusion}

In a general way, dealing with basic belief assignments on large
frames of discernment need a proper encoding of the focal sets. We
propose to use hashtables for this purpose, but this not the only way.
We propose two categories of methods for restricting any bba to a bba
modest in focal set size.

We extend the principle of isopignistic to other bodies of evidence to
build a bba with only $2n\!-\!1$ focal elements, respecting both the
pignistic probability and another body of evidence of the original
bba. We first determine the value of the bodies of evidence on some
simple elements, and then determine the restricted focal set. A linear
equation gives the restricted bba.

Trying to restrict the focal set to a number of respresentative
elements leads to a NP-Complete problem. We adapt the $k$-mean
algorithm to build a heuristical solution. It is more expensive, but
it does not need to define {\em a priori} a focal set, and can adapt
to more situations.

\bibliography{Belief12-Osswald}

\end{document}